# SPECTROSCOPY IN EXTREMELY THIN VAPOR CELLS : SENSITIVITY ISSUES

**M. Ducloy and D. Bloch**

*Laboratoire de Physique des Lasers, CNRS-Université Paris13,
99 Av. J.B. Clément, 93430 Villetaneuse, France*

Since a few years, we have started the spectroscopic analysis of Extremely Thin Cell (ETC) of dilute vapor. The anisotropy intrinsic to such cells justifies the partial or total elimination of Doppler broadening [1-5] under normal incidence, owing to the enhanced transient response of the atoms with the longest free paths from wall-to-wall. One of our present purpose is to develop an effective method for the evaluation of the atom-surface interaction. This communication focuses on sensitivity issues - a long-time concern of Jan Hall[6]- in this peculiar type of spectroscopy. With these small and often submicrometric slices of vapor, the most uncommon features are the relatively small number of interacting atoms, and the fact that essential results are already obtained in the frame of *linear* spectroscopy.

The fabrication of ETCs of alkali-metal vapors is achieved by contacting two thick windows with a very thin sapphire spacer [2]. The thickness varies locally, within a typical range 20-1000 nm, as a result of the stress induced onto the windows by the atmospheric pressure. Spectroscopy in these ETCs has been mostly developed through transmission measurements [3], but fluorescence[2,4], reflection spectroscopy [5], and Faraday rotation have been also demonstrated. The major results until now were obtained with a resonant 1-photon excitation, although 2-photon schemes have also been demonstrated.

Classically, the shot-noise sets the ultimate limit to sensitivity. In the principle, this makes the direct detection of emitted particles, like fluorescence or ionization, preferable to absorption detection, as long as the emission signal is obtained on a zero background. This may seem exemplified with the fluorescence detection of an ETC, with a sub-Doppler resonance in a Cs or Rb sub-micrometer cell possibly detected at room-temperature (the current absorption experiments in an ETC typically require a 10-100 times higher density). Actually, only a small solid angle of the emitted fluorescence is detected (*i.e.* reducing the number of observed events), and the light scattering generates a nonzero background at the same wavelength. Specific features of the ETC can also restrict the effective applicability of a fluorescence detection: when multiple absorption/emission cycles occur, the fluorescence signal may partly originate from those non-irradiated regions, whose thickness has not been measured. Also, the narrower fluorescence lineshape (relatively to absorption [4]) is actually a consequence of the selection of those rare atoms that experience, after excitation, a free flight long enough to get de-excited in a radiative process, instead of decaying through a quenching collision with the surface.

In a linear absorption process, the ultimate sensitivity depends on the ratio between the shot-noise of the beam falling onto the detector, and the signal field, proportional to the incident field itself. Practically, the beam has to be focused (spot size $\leq 0.01$ mm²) to keep the local thickness of the ETC reasonably constant, and a linear absorption experiment on the Cs $D_1$ or $D_2$ line is conducted with $P \leq 1$ µW (~$5.10^{12}$ photon/s) to avoid saturation. One extrapolates a theoretical minimal detectable absorption ~$10^{-6}$ for a typical 0.3 s integration time. The intrinsically small amplitude noise of laser diodes enables one to flirt in the experiments with this theoretical value. If a stronger power could be used, the sensitivity would theoretically increase, and the detectivity would fall in the principle well below $10^{-6}$, a range of values hardly achieved owing to the limited linearity of detectors.

This obstacle to an increased sensitivity can be addressed by a modified detection scheme, with a specific decrease of the intensity of the detected field, responsible for the shot-noise, while keeping the weak signal field unchanged. This is the basis for optical homodyne detection, with an adjustable *local oscillator* (LO), combined with the signal field, yielding a signal $|E_{LO}|^2 - |E_{LO}+E_{sig}|^2 \approx 2\ (E_{LO})^*E_{sig}$. In laser spectroscopy, this approach can be at least traced back to polarization spectroscopy [7], with the LO field reduced to the desired level by a nearly crossed-polarizer. An optimal sensitivity is hence reached when the LO amplitude is comparable with the signal field, in order not to add to the signal shot-noise. Extensions of this approach include Faraday rotation spectroscopy (also with nearly crossed-polarizers), or selective reflection spectroscopy on a nearly zero background (*e.g.* at the Brewster angle[8], or at the interface with a Fabry-Perot-like window[9]), along with reflection detection on an ETC when its thickness is chosen equal to $\lambda/2$ [10]. However, in all these schemes, the LO field propagating parallel to the signal field is never perfectly extinct, bringing this inconvenient that the lineshapes are a combination of the real and imaginary parts of the signal field amplitude, that vary with the LO field attenuation. To our knowledge, only nonlinear wave-mixing, with its additional spatial modulation [11], has allowed for an effective nulling of the LO field, with observation of those special lineshapes given by $|E_{sig}|^2$, corresponding to a background-free detection of all absorption events. However, it is not valid for linear spectroscopy, and hence of a limited interest for ETC spectroscopy.

Optical heterodyne spectroscopy [12,6], developed at the heroic era of noisy dye lasers, is a high-frequency version of the homodyne technique, where an applied FM or AM allows for relaxation time measurements. With diode lasers, the sensitivity is not truly improved by a fast modulation. However, the FM technique, through its natural frequency-derivation of the standard lineshape, offers the remarkable advantage [10] of turning a sub-Doppler linear transmission in an ETC into a genuine Doppler-free signal, in the same well-known way as in selective reflection. Indeed, the linear atomic transient response combines an absorptive component with a broad-wing *dispersive* response, that is not strictly-speaking "velocity-selective". In the direct detection, the velocity integration converges only because of the finite wings of the Maxwell distribution (*i.e.* fast atoms contribute also to the signal), while the FM derivation ensures convergence even for an infinite Doppler width.

These discussions on sensitivity may look academic for such a robust system as an alkali-metal vapor. They are actually an essential issue for the envisioned application of this linear Doppler-free spectroscopy to those weak molecular lines that remain hard to saturate [13].